\documentclass{sf2a-conf2018}
\usepackage{graphicx}
\usepackage{hyperref}
\usepackage[]{natbib}  
\usepackage{epstopdf}

\def\BibTeX{{\rm B\kern-.05em{\sc i\kern-.025em b}\kern-.08em
    T\kern-.1667em\lower.7ex\hbox{E}\kern-.125emX}}
\bibpunct{(}{)}{;}{a}{}{,}  

\newcommand{\txs}{TXS\,0506$+$056}
\newcommand{\gray}{$\gamma$-ray}

\begin{document}

\TitreGlobal{SF2A 2018}

\title{\vspace{-2.2cm}
Gammas and neutrinos from \txs}

\runningtitle{Gammas and neutrinos from \txs}

\author{M. Cerruti$^{1,}$}\address{Sorbonne Universit\'e, Universit\'e Paris Diderot, Sorbonne Paris Cit\'e, CNRS/IN2P3, Laboratoire de Physique Nucl\'eaire et de Hautes Energies, LPNHE, 4 Place Jussieu, F75252 Paris, France}
\address{Institut de Ciències del Cosmos (ICCUB), Universitat de Barcelona (IEEC-UB), Mart\'i i Franqu\`es 1, E08028
Barcelona, Spain\\
matteo.cerruti@icc.ub.edu}
\author{A. Zech}\address{LUTH, Observatoire de Paris, PSL Research University, CNRS, Universit\'e Paris Diderot, 5 Place Jules Janssen, 92190 Meudon, France}

\author{C. Boisson$^3$}

\author{G. Emery$^1$}

\author{S. Inoue}\address{RIKEN, Institute of Physical and Chemical Research, 2-1 Hirosawa, Wako, Saitama, 351-0198, Japan}

\author{J.-P. Lenain$^1$}




\setcounter{page}{237}


\maketitle


\begin{abstract}
While blazars have long been one of the candidates in the search for the origin of ultra-high energy cosmic rays and astrophysical neutrinos, the BL Lac object \txs\ is the first extragalactic source 
that is correlated with some confidence with a high-energy $\nu$ event recorded with IceCube. 
At the time of the IceCube event, the source was found in a high state in \gray s with Fermi-LAT and MAGIC. 
We have explored in detail the parameter space of a 
lepto-hadronic radiative model, assuming a single emitting region inside the relativistic jet. We 
present the complete range of possible solutions for the physical conditions of the emitting region and its 
particle population. For each solution we compute the expected $\nu$ rate, and discuss the impact of this event on our general understanding of emission processes in blazars.\end{abstract}


\section{Introduction}
\citet{0506science} reported the detection of the high-energy neutrino IceCube-170922A, for the first time coinciding spatially and temporally with a blazar in an elevated \gray\ flux state, as observed with {\it Fermi}-LAT and MAGIC. A chance correlation is rejected at the 3$\sigma$ level. The blazar in question, \txs, is a BL Lac object and its redshift was measured to be $z=0.337$ \citep{paiano}. While the probability of 56.5$\%$ for this single $\nu$ to be truly astrophysical does not yet firmly establish blazars as sources of high-energy $\nu$s, this detection represents the first direct observational indication for such a link. The simplest scenario that may explain correlated electromagnetic and neutrino emission in blazars is the one-zone, lepto-hadronic model, where a magnetized compact region inside the relativistic jet carries a population of relativistic electrons and protons. Neutrinos are generated as part of the pion-decay chain in p-$\gamma$ interactions, while synchrotron-pair cascades of secondary particles and/or proton-synchrotron radiation are responsible for the high-energy part of the electromagnetic spectral energy distribution (SED), the low-energy part being usually ascribed to synchrotron radiation from primary electrons.
In this contribution, we present the results of a systematic hadronic modeling of \txs, including solutions where the \gray\ component is dominated by proton synchrotron radiation, or synchrotron-self-Compton (SSC) radiation, with a sub-dominant hadronic component. We constrain the parameter space from the electromagnetic observations, and for each solution we compute the expected $\nu$-rate. Assuming that the association between \txs\ and IceCube-170922 is genuine, we then discuss the impact of this event on blazar emission models. For further details, see \citet{leha0506}.   
\section{Numerical simulations}

\begin{figure}[ht!]
 \centering
 \includegraphics[width=0.4\textwidth,clip]{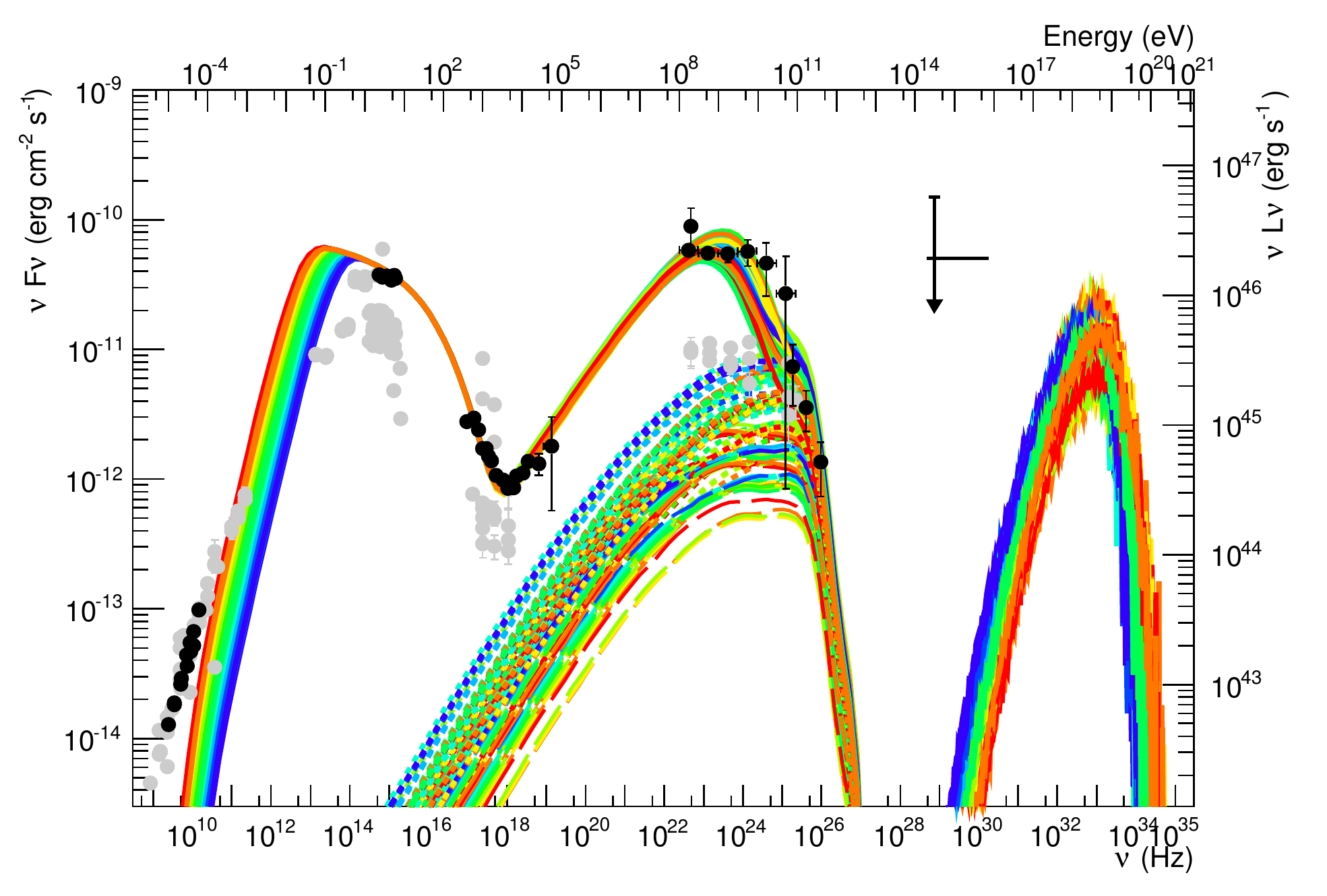}%
 \includegraphics[width=0.4\textwidth,clip]{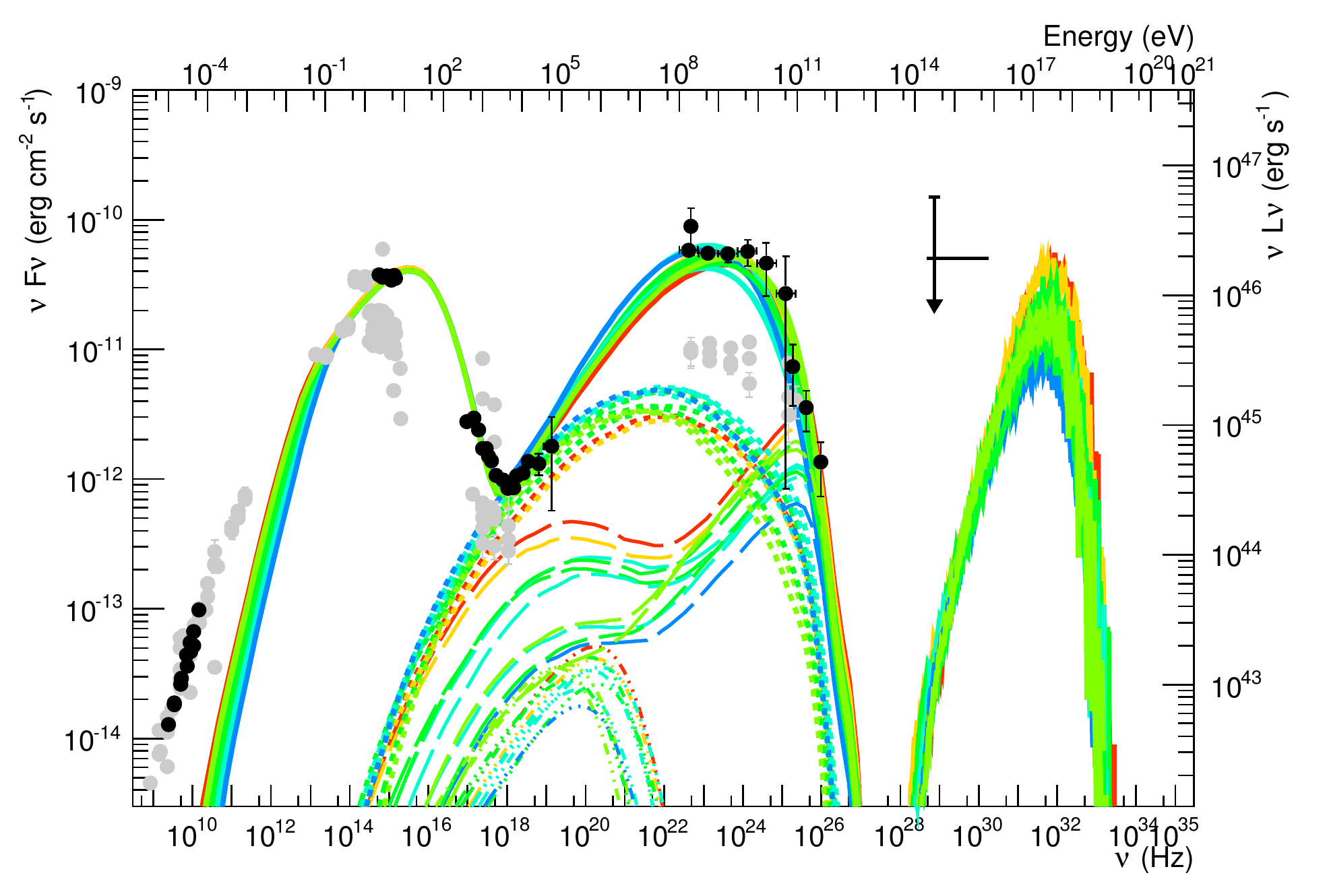}      
  \caption{{\bf Left:} Proton synchrotron solutions. {\bf Right:} Mixed lepto-hadronic solutions }
  \label{myfig}
\end{figure}

The \textit{LeHa} code \citep{lehauhbl} is used to simulate electromagnetic and neutrino emission from \txs. It has been developed to describe the stationary \gray\ emission from BL Lacertae objects, taking into account all relevant leptonic and hadronic radiative processes. 
The number of free parameters of the model is 15: 3 for the emitting region (the Doppler factor $\delta$, the magnetic field $B$, and the radius $R$), 12 for the primary population of leptons and protons (the 4 indices of the broken power-law distributions, $\alpha_{e/p , 1/2}$; the minimum, break, and maximum Lorentz factors $\gamma_{e/p,\textnormal{min/break/max}}$; and the normalizations $K_{e/p}$). We reduce the number of free parameters to 8, by assuming that protons and electrons are co-accelerated (and thus share the same injection index), that the maximum energy of protons is constrained by equating acceleration and cooling time-scales, and that $\gamma_{e,\textnormal{break}}$ and $\alpha_{e,2}$ are determined by cooling. $\gamma_{e,\textnormal{max}}$  and $K_e$ are adjusted to fit the low-energy part of the SED, the index of the primary particles is fixed to $2.0$, while the other 5 parameters ($\delta$, $B$, $R$, $K_p$ and $\eta$ which is the efficiency factor of the proton acceleration term) are systematically scanned. For each model we compute the $\chi^2$ with respect to the observations, and we select solutions with $\Delta \chi^2$ within $\pm 1\sigma$. \\
The first scenario (left plot of Fig. \ref{myfig}) is a proton synchrotron one, in which the emission from hard-X-rays to \gray s is due to proton synchrotron with a pion cascade component emerging at VHE. In this case $\delta = 35-50$, $B=(0.8-32)$ G, $R = 1\times10^{15}-9.7\times10^{16}$ cm, $\gamma_{p,\textnormal{max}} = 4\times10^8 - 2.5\times 10^9$ and the total jet power is $L = 8\times 10^{45}-1.7\times10^{48}$ erg s$^{-1}$. The expected neutrino rate in this scenario is $0.006-0.16$ yr$^{-1}$ in the full IceCube energy band, which drops to $2.4\times10^{-5}-0.002$ if we consider only the 0.183 TeV - 4.3 PeV energy band of IceCube 170922A.\\
The second scenario (right plot of Fig. \ref{myfig}) is a mixed lepto-hadronic one, in which the SSC emission dominates the high-energy SED component, with a hadronic emission that emerges in hard-X-rays (as Bethe-Heitler cascade) and VHE (as pion cascade). In this case the parameters are $\delta = 30-50$, $B=(0.13-0.65)$ G, $R = 2\times10^{15}-1.5\times10^{16}$ cm, $\gamma_{p,\textnormal{max}} = 6\times10^7 - 2\times 10^8$ and the total jet power is $L = 3.5\times 10^{47}-3.5\times10^{48}$ erg s$^{-1}$. The expected neutrino rate in this scenario is $0.11-3.0$ yr$^{-1}$ in the full IceCube energy band, and $0.008-0.11$ if we consider only the energy band of IceCube 170922A. In this scenario, given the high neutrino flux, it is important to estimate also the neutrino rate with the IceCube effective area for point-like sources, which results in $0.3- 6.9$ yr$^{-1}$. The solutions with the highest neutrino rates face thus a difficulty in explaining why only one neutrino has been observed. It is important to underline here that the neutrino rate does depend on the assumed efficiency of the acceleration time-scales of protons.
\section{Conclusions}
We performed an extensive study of the hadronic model parameter space for \txs, identifying the solutions which fit the SED and computing the expected neutrino rates. The first result is that, while proton synchrotron solutions can correctly reproduce the electromagnetic emission, they do not produce enough neutrinos, and, if the association of \txs\ and IceCube 170922A is genuine, they are strongly disfavoured. Lepto-hadronic solutions, on the other hand, can accout for this neutrino event, even though they are quite demanding in terms of jet power. In addition, they are constrained by the non-detection of PeV neutrinos with the IceCube point-like search algorithm. For more thorough explanations on this study, the reader is refered to \citet{leha0506}.

\begin{acknowledgements}
This work is supported by JSPS KAKENHI Grant Number
JP17K05460 for SI. MC has received financial support through the Postdoctoral Junior Leader Fellowship Programme from “la Caixa” Banking Foundation
\end{acknowledgements}

\bibliographystyle{aa}  
\bibliography{sf2a-template} 

\end{document}